\documentclass[12pt,preprint]{aastex}
\usepackage{epsfig}
\usepackage{natbib}
\usepackage{lscape}
\makeatletter
\newcommand*{\rom}[1]{\expandafter\@slowromancap\romannumeral #1@}
\makeatother
  \usepackage{hyperref}
\usepackage{amsmath} 

  \usepackage{courier}
\usepackage{graphicx}

\shorttitle{IUE Be survey}
\shortauthors{Wang et al.}

\begin{document}

\received{}
\accepted{}
\setcounter{footnote}{0}

\title{Detection of additional Be+sdO systems from \emph{IUE} spectroscopy}

\author{Luqian Wang, Douglas R. Gies} 
\affil{Center for High Angular Resolution Astronomy and  
 Department of Physics and Astronomy,\\ 
 Georgia State University, P. O. Box 5060, Atlanta, GA 30302-5060, USA} 
\email{lwang@chara.gsu.edu, gies@chara.gsu.edu} 
 
\author{Geraldine J. Peters\altaffilmark{1}} 
\affil{Space Sciences Center, University of Southern California, Los Angeles, CA 90089-1341, USA}  
\email{gpeters@usc.edu} 
 
\altaffiltext{1}{Guest Observer with the {\it International Ultraviolet Explorer}.} 
 
\slugcomment{} 
\paperid{}

%%%%%%%%%%Abstract%%%%%%%%%%%%%%%%%%%%%%%%%%%%%%%%%%%%%
\begin{abstract}
There is growing evidence that some Be stars were spun up through mass transfer in a 
close binary system, leaving the former mass donor star as a hot, stripped-down object. 
There are five known cases of Be stars with hot subdwarf (sdO) companions that were 
discovered through \emph{International Ultraviolet Explorer (IUE)} spectroscopy. 
Here we expand the search for Be+sdO candidates using archival FUV spectra from \emph{IUE}. 
We collected \emph{IUE} spectra for 264 stars and formed cross-correlation functions (CCFs) 
with a model spectrum for a hot subdwarf.  Twelve new candidate Be+sdO systems were found, 
and eight of these display radial velocity variations associated with orbital motion. 
The new plus known Be+sdO systems have Be stars with spectral subtypes of B0 to B3, and the 
lack of later-type systems is surprising given the large number of cooler B-stars in our sample.
We discuss explanations for the observed number and spectral type distribution of
the Be+sdO systems, and we argue that there are probably many Be systems with 
stripped companions that are too faint for detection through our analysis. 
\end{abstract}

 
\setcounter{footnote}{1}

%%%%%%%%%%%% Section 1 %%%%%%%%%%%%%%%%%%%%%%%%%%%

\section{Introduction} 
Be stars are rapidly rotating B-type, non-supergiant, stars that show or 
have shown H$\alpha$ emission in their spectra. Their rotational velocities 
may reach more than 75$\%$ of the critical velocity \citep{slettebak1966,rivinius2013}. 
\citet{pols1991} suggested that their rapid rotation results from past 
mass transfer in a close binary system. The initially more massive star 
expands away from main-sequence stage after hydrogen core exhaustion and 
fills its Roche-lobe. Mass transfer from the evolved massive star to the 
less massive gainer star causes the latter to spin up due to conservation 
of angular momentum. The orbit shrinks until the two stars reach comparable masses, 
and then subsequent mass transfer causes the orbit to expand. This continues until 
the donor star loses its outer envelope and the core obtains a size smaller 
than the Roche-lobe. If the donor star is left with a mass above the 
Chandrasekhar limit, then the stars will evolve into an X-ray binary system, 
consisting of a Be star and a neutron star or black hole.  Lower mass donor remnants 
become a faint, hot, subdwarf (sdO) or a white dwarf (WD).

Such sdO and WD stars are difficult to detect because they are usually lost 
in the glare of their massive companions and their small mass creates only a 
small orbital reflex motion in the Be star. The subdwarfs are hot, so it is best 
to search for this type of remnant in the far ultraviolet because they contribute 
relatively more flux there and their spectra are rich in highly ionized metallic lines. 
Such FUV spectroscopy investigations have led to the detection of a subdwarf companion 
in five systems. \citet{thaller1995} discovered the hot subdwarf companion of 
$\phi$~Per using a Doppler tomography algorithm, which uses the radial velocities 
of the components to reconstruct their individual spectra. The highly ionized lines 
of the sdO star were clearly visible in the reconstructed secondary spectrum based 
on 16 Short Wavelength Prime (SWP) camera, high dispersion (H) spectra obtained with 
the \emph{International Ultraviolet Explorer}. This discovery was confirmed by 
\citet{gies1998} through \emph{Hubble Space Telescope} spectroscopy, and their study 
suggested that the subdwarf companion has $T_{\rm eff} = 53 \pm 3$ kK and 
$\log{g} = 4.2 \pm 0.1$. \citet{peters2008} combined optical spectra and 96 
\emph{IUE} spectra to confirm the binarity of FY~CMa, and they detected the 
hot companion through analysis of cross-correlation functions (CCFs) of the UV spectra 
with that for a hot stellar template. Their study indicated that the secondary 
has $T_{\rm eff} = 45 \pm 5$ kK and $\log{g} = 4.3 \pm 0.6$. A similar analysis was 
done by \citet{peters2013} for 59~Cyg, and they reported that the detected companion 
has $T_{\rm eff} = 52.1 \pm 4.8$ kK and $\log{g} = 5.0 \pm 1.0$ based on a large set 
of 157 \emph{IUE} spectra. Subsequently, \citet{wang2017} analyzed 23 \emph{IUE} 
spectra of 60~Cyg through CCFs with a hot stellar template, and they estimated that 
the hot subdwarf companion has $T_{\rm eff} = 42 \pm 4$ kK. The fifth detection was 
made by \citet{peters2016} using 88 \emph{IUE} spectra to detect a faint signal of 
a hot companion in HR~2142, which indicated that the companion has 
$T_{\rm eff} \ge 43 \pm 5$ kK.  

The detection of the subdwarf companions of the confirmed systems benefited from 
the large number of observations available in the \emph{IUE} archive in order 
to take advantage of the $\sqrt{n}$ improvement in S/N by combining all the 
observations in the analysis. However, relatively bright subdwarf companions 
should be detected even in a single \emph{IUE} observation through calculations 
of CCFs with a model spectrum for a hot star. The goal of this work is to detect 
other Be+sdO systems by searching for such relatively bright companions through 
analysis of individual spectra for a large sample of Be stars. 

Our survey will help identify binary systems with subdwarf companions for 
future follow-up studies to determine the orbital and stellar parameters that 
are needed for critical comparisons with models for the evolution of 
stripped cores \citep{althaus2013}. This survey is important for studies of 
massive star evolution, since a large fraction of massive stars have nearby 
stellar companions \citep{sana2012} and binary interactions play a key role 
in their destinies \citep{demink2014}. Hot evolved companions may contribute 
significantly to the UV flux of stellar populations \citep{han2010} and 
constitute a missing contribution to spectral synthesis models \citep{bruzual2003}. 
Massive helium star remnants probably explode as hydrogen deficient supernovae 
(SN \rom{1}b and SN \rom{1}c; \citealt{eldridge2013}), so a determination of 
their numbers and properties is closely related to the numbers and kinds of 
core collapse SN we observe. The rotational properties of SN progenitors 
dictate the spins of their neutron star and black hole remnants, and fast rotation 
may be responsible for the long duration $\gamma$-ray bursters that form in the 
core collapse of massive stars \citep{cantiello2007}. 

Here we present the results of the survey for sdO companions among rapidly rotating 
hot stars from an analysis of \emph{IUE} spectra. Section 2 presents our subdwarf 
flux search method that is based on a cross-correlation analysis of the UV spectra 
with a model spectral template. We discuss the 12 new candidate Be+sdO systems in Section 3. 
Our final results and their consequences are summarized in Section 4.

%%%%%%%%%%%% Section 2 %%%%%%%%%%%%%%%%%%%%%%%%%%%

\section{Search for sdO companions}

% Data selection
The main sample of Be stars was adopted from the list of \citet{yudin2001}, 
who presented an analysis of intrinsic polarization, projected rotational velocity, 
and IR excess of 627 Be stars. We combed through the \emph{IUE} archive to find 
these targets, and we collected 3092 SWP/H spectra of 226 stars from the archive. 
We further expanded the sample by adding 111 SWP/H spectra of 38 rapidly rotating, 
non-emission stars with projected rotational velocity $V\sin{i} > 300$ km s$^{-1}$.

% Data reduction
We downloaded the SWP/H spectra of these 264 stars from 
MAST\footnote{https://archive.stsci.edu/iue/}. The spectra were reduced and 
rectified following the procedures reported in \citet{wang2017}, 
except that we left the interstellar medium lines in place. 
% CCF construction
The estimated temperatures of the five known sdO companions all have 
$T_{\rm eff} > 40$ kK. Thus, we adopted a synthetic spectrum with $T_{\rm eff} = 45$ kK 
from the grid of \citet{lanz2003} that we used earlier \citep{wang2017}, 
and we cross-correlated it with all the observed spectra. 
We excluded the beginning and ending regions and very broad or blended 
line regions (including the \ion{Si}{3} $\lambda1300$ complex, 
\ion{Si}{4} $\lambda\lambda1394,1403$, \ion{Si}{2} $\lambda\lambda1527,1533$, 
and \ion{C}{4} $\lambda1550$) that were replaced by a linear interpolation 
across the adjacent continuum to avoid introduction of broad features into the CCFs.

% Target selection
We began our search for hot companions by forming the ratio of the CCF 
maximum height (peak signal S) within a velocity range of $\pm200$ km s$^{-1}$ 
(larger than the typical span of Doppler shift of the known binaries) 
to the standard deviation of the CCF in higher velocity portions (background noise N), 
and we then calculated the average peak-to-background ratio (S/N) from the individual
ratios for all the available spectra for each star. The average S/N ratios of 
the known Be+sdO systems ($\phi$~Per, FY~CMa, 59~Cyg, and 60~Cyg) all have S/N  $>3.0$. 
Thus, we set this CCF peak-to-background ratio as the lower limit to select 
candidate sdO binaries. Null detections of companion stars with CCF S/N ratios below 
the selection criterion are listed in Table~1. If the stars in Table~1 host 
hot companions, then their sdO components must be too faint to detect in 
individual \emph{IUE} spectra. Table~1 includes HR~2142 (=HD~41335), which only displays 
the weak signal of the companion in a subset of spectra \citep{peters2016}. 
On the other hand, a strong signal appeared and met the selection criterion for 66 stars, 
forming a preliminary list of potential binary systems with relatively bright subdwarf companions. 
We collected the spectral classifications and projected rotational velocities of these targets 
from the literature. The HD number, star name, HIP number, spectral classification, 
projected rotational velocity, number of SWP/H spectra available in the \emph{IUE} archive, 
average CCF S/N ratio, and references for the spectral types and $V\sin{i}$ are tabulated 
in Table~2.  
             
% Candidate selection criteria
We adopted detection criteria for Be+sdO binaries that are based upon the characteristics 
of the five known systems. The best known system (and the one with the highest CCF S/N 
ratio in our sample) is $\phi$~Per (HD~10516), and we show in Figure~1, top left panel, 
two CCFs for $\phi$~Per at different orbital velocity extrema. The CCF peaks are narrow 
(indicative of a small projected rotational velocity) and show large velocity excursions 
due to the large semi-amplitude associated with the low mass sdO star \citep{gies1998}. 
All five known Be+sdO binaries share these properties, so we adopted two detection 
criteria based upon them.

% First criterion
The first criterion is that any CCF signal from the sdO component must be narrow 
with a half width at half maximum (HWHM) much smaller than the projected rotational 
broadening $V \sin i$ associated with the primary Be star target. 
Most Be stars are rapid rotators with large $V \sin i$, so their associated CCFs 
are very broad (see the case of $\gamma$~Cas = HD~5394 in the lower left panel of Fig.~1), 
although a few pole-on Be stars do show smaller projected rotational broadening 
(see the CCFs of HD~120991 in the lower, right panel of Fig.~1). 
Note that this assumption biases the results against detection of rapidly rotating 
sdO components, but we doubt such exist in general because their progenitors probably 
became synchronous rotators in long period orbits during the earlier mass transfer phase. 
Note that the CCF signal from correlation with the Be star spectral features will 
become larger for hotter Be stars, so that the CCF will be more dominated by the 
Be star signal in earlier-type targets. We show, for example, the cases of CCFs for 
a hot (HD~155806) and a mid-temperature (HD~174237) emission-line star in the middle 
left and right panels, respectively, of Figure~1. The sensitivity of the CCFs 
to the temperature of the Be star means that it will become progressively more 
difficult to distinguish the signal from a sdO component from that of the Be star 
at high temperature, so our methods may be biased against detection of sdO companions 
among the hotter Be stars. Nevertheless, it is possible to discern the narrow sdO 
component against broader Be component in some cases of hotter Be stars 
(see the case of FY~CMa = HD~53978 in the upper right panel of Fig.~1, 
which shows the narrow peak of the sdO star atop the broader signal from the Be star). 

% Second criterion
The second criterion is based upon the expected orbital motion of the sdO component.  
Post-mass transfer binaries are expected to have extreme mass ratios, so that the 
orbital semiamplitudes of the sdO components will be large ($\approx 50 - 100$ km~s$^{-1}$) 
compared to those of the Be stars ($<10$ km~s$^{-1}$) for binaries with periods 
of a few months or less.  Thus, our second criterion for sdO candidates is that 
their CCF signals should show a velocity range greater than $1/2$ of the HWHM of 
the CCF peak. This criterion tacitly assumes that enough spectra exist to sample 
the full range of orbital motion, but this cannot be fulfilled in cases where only 
a few spectra are available. Note that application of this criterion will impose a 
bias against detection of binaries in low inclination and long period orbits 
(with small orbital semiamplitude). 

% Results
The results of applying these two detection criteria are summarized in a code in 
the last column of Table~2. We find that 50 of the 66 targets in the list of 
large S/N cases are best explained as the result of correlation with the Be star 
spectrum itself (indicated by a ``P'' for primary in Table~2). 
These are generally hotter Be stars in which the CCF signal has a half-width 
comparable to the published $V \sin i$, and their CCF peaks show little evidence 
of significant orbital Doppler shifts. Some of these cases are discussed in the Appendix. 
Next, the application of the two criteria led to the successful re-identification 
of four known Be+sdO binaries ($\phi$~Per, FY~CMa, 59~Cyg, and 60~Cyg), and these 
are indicated by an ``S'' (for subdwarf) in Table~2. The criteria also led to the 
detection of eight additional candidate Be+sdO binaries that are labeled with a ``C'' in Table~2. 
Finally, we list four systems in which a strong and narrow CCF peak was found, 
but only one or two spectra are available in the {\it IUE} archive so that 
we could not apply the second criterion of velocity variability. 
These four potential candidates are indicated by a ``C?'' code in Table~2. 
Radial velocities from Gaussian fits of the upper 80\% of the CCF peaks are listed 
in Table~3 for these eight candidate and four potential candidate systems, 
but measurements from noisy spectra with weak CCF signals were omitted. 
The number and time distribution of these measurements are insufficient 
to find orbital periods and other elements, but they are included here for 
future use once orbits are determined (perhaps by ground-based spectroscopy 
of the Be components). All these new detections are discussed further in the next section. 

%%%%%%%%%%%% Section 3 %%%%%%%%%%%%%%%%%%%%%%%%%%%

\section{Candidate Be+sdO systems}

We identified twelve subdwarf candidates from the CCF analysis that display a 
narrow peak as expected from correlation with the hot subdwarf spectrum template. 
We observe significant Doppler shift variations in the CCF signals of eight systems 
due to the orbital motion of the subdwarf companion. Figure~2 shows the apparent CCFs 
of the eight candidate systems for spectra observed near the velocity extrema (see Table~3). 
The remaining four potential subdwarf candidates in the sample display a narrow peak, 
but little or no information about their Doppler shifts is available due to the limited 
number of spectra available. The CCFs for these four cases are shown in Figure~3. 
The peak height of the CCF scales approximately with the subdwarf flux contribution 
if the model template is a reasonable match \citep{wang2017}, so the new candidates 
contribute about $2-5\%$ of the FUV monochromatic flux (and probably less at optical 
wavelengths). 

None of the candidates are known close binaries, but all are worthy of follow-up investigation. 
We summarize a brief literature review for each candidate below. Many of the candidate Be+sdO 
systems have additional companions detected through speckle interferometry or optical imaging, 
but their orbital periods must be decades or longer. Consequently these visually resolved companions 
are unrelated to the subdwarf companions that are the remnants of interactions in close binaries.  
The speckle interferometry observations reveal companions in the angular separation range of
$0.035 < \rho < 1\farcs5$ and brighter than $\triangle m < 3.0$ mag \citep{mason1997}. 

\noindent
% HD 29441
{\bf{HD 29441 (V1150 Tau; B2.5 \rom{5}ne; S/N = 3.77)}}. Based on measurements of 
six optical spectra, \citet{chini2012} found that the star is radial velocity constant, 
perhaps indicating that the sdO is a low mass object or that the orbit is very long. \newline 
% HD 43544
{\bf{HD 43544 (B3 \rom{5}; S/N = 5.51)}}. \citet{huang2010} argued that the star had a 
significant radial velocity shift between measurements of two optical spectra, 
indicative of the orbital motion of the Be star. No other companions have been detected 
through speckle interferometry from \citet{mason1997}. \newline
% HD 51354
{\bf{HD 51354 (QY Gem; B3 \rom{5}e; S/N = 5.39)}}. \citet{chojnowski2017} observed only a small
scatter of 3.0 km~s$^{-1}$ in six radial velocity measurements from the APOGEE survey. 
% HD 60855
{\bf{HD 60855 (V378 Pup; B3 \rom{4}; S/N = 4.98)}}. The CCFs of the star display a narrow 
central peak that sits atop a broader signal from the Be star. The measured velocities from 
the CCFs of the six available spectra did not display large velocity shifts, perhaps 
indicating that the sdO is in a long period, slow moving orbit. However, \citet{huang2006} 
found a significant velocity shift between their two spectra indicative of possible orbital 
motion of the Be star. \citet{mason1997} found no evidence from speckle interferometry for 
another companion in the separation range of 0.04 to 1 arcsec. \citet{mason2001} note that 
this star is a member of the NGC 2422 cluster, and the nearest companion has separation of 
5.3 arcsec from the star. \newline
% HD 113120
{\bf{HD 113120 (LS Mus; B2 \rom{4}ne; S/N = 4.35)}}. Based on spectroscopic measurements from 
five optical spectra, \citet{chini2012} found that the star is radial velocity constant, which 
suggests a small orbital semi-amplitude for the Be star. Nevertheless, we observe relatively 
large Doppler shifts for the sdO component. \citet{hartkopf1996} detected a companion with 
angular separation of 0.557 arcsec from the star through speckle interferometry. \newline
% HD 137387  
{\bf{HD 137387 ($\kappa^1$ Aps; B2 \rom{5}npe; S/N = 5.21)}}. \citet{lindroos1985} reported 
that the star belongs to a visual binary system with a companion of estimated spectral type 
of K7 \rom{4} and a projected separation of 1470 AU from the star. \newline
% HD 152478
{\bf{HD 152478 (V846 Ara; B3 \rom{5}npe; S/N = 3.69)}}. \citet{hoogerwerf2001} mention this object as a 
possible runaway star.  \citet{bruijne2012} list a radial velocity of 19 km~s$^{-1}$. \newline
% HD 157042
{\bf{HD 157042 ($\iota$ Ara; B2.5 \rom{4}e; S/N = 4.09)}}. Based on optical studies, \citet{lindroos1985} 
reported that the star has a companion with estimated spectral type of G5 \rom{3}-\rom{4} and a 
separation of 42.8 arcsec from the primary component. \newline
% HD 157832
{\bf{HD 157832 (V750 Ara; B1.5 \rom{5}e; S/N = 4.21)}}. Based on observations from XMM-Newton and 
optical spectroscopy, \citet{oliveira2011} classified the star as the coolest $\gamma$ Cas analogue. 
Based on the presence of \ion{Fe}{21} and \ion{Fe}{22} recombination lines and fluorescence features, 
\citet{garcia2015} confirmed the X-ray properties of the star and classified it as a component of high-mass X-ray binary. \newline 
% HD 191610 
{\bf{HD 191610 (28 Cyg; B3 \rom{4}e; S/N = 3.02)}}. \citet{abt1978} reported that the star is  
radial velocity constant from spectroscopic studies based on 25 optical spectra. Based on an analysis 
of space photometry and H$\alpha$ line profiles, \citet{baade2017} concluded that the large-amplitude 
frequencies due to multiple non-radial pulsation modes are responsible for the observed short- and 
medium-term variability, and these pulsation modes are also related to the modulation of mass transfer 
events from the star to the disk of 28 Cyg. No other companions have been found by \citet{mason1997} 
through speckle interferometry. \newline
% HD 194335
{\bf{HD 194335 (V2119 Cyg; B2 \rom{3}e; S/N = 4.54)}}. The star is listed as a shell star by \citet{hoffleit1991}. 
\citet{plaskett1920} suggested that the object is a possible spectroscopic binary. \newline
% HD 214168
{\bf{HD 214168 (8 Lac B; B1 \rom{4}e; S/N = 3.03)}}. \citet{hoffleit1991} also identified this star 
as a shell star. It is a member of the Lac OB1 association. \citet{mason1997} identified a companion 
with an angular separation of 0.042 arcsec using speckle interferometry. \citet{shatskii1998} and 
\citet{mason2007} confirmed that the star forms a double system with HD 214167 (B1.5 \rom{5}s) with 
an angular separation of 22.24 arcsec.

%%%%%%%%%%%% Section 4 %%%%%%%%%%%%%%%%%%%%%%%%%%%

\section{Conclusions}

We identified eight subdwarf candidates and four potential candidates, 
and by including the five Be+sdO systems known from previous studies, 
this leads to a total of 17 detections in the sample of 264 stars, 
for a detection rate of $6\%$.  The CCF S/N ratios have a range between 
3.0 and 5.5 with a mean of 4.3, and these ratios are generally less than 
or comparable to those of the known systems.  The distributions 
with spectral type of the known plus candidate targets and of the full 
sample are shown in Figure~4.  Most of the new candidates have spectral 
types of B2-B3, which are relatively cooler compared to the primaries of 
the known Be+sdO systems with spectral types of B0.5-B1.5.  However, 
we found no companions among the cooler, later-type B-stars, despite the 
fact that such hot companions should be more readily detected as 
they dominate more of the FUV flux distribution relative to cooler,
main sequence companions.  There are 109 targets with spectral types B4 
and later in our sample ($41\%$).  If the Be+sdO systems had the same 
spectral type distribution as the whole sample, then we would expect to
have found 7 of 17 systems among the B4-A0 group, yet none were detected. 

Binary star population models make different predictions about the numbers
of expected Be+He star systems (which we may compare to the observed Be+sdO systems)
as a function of Be star mass or spectral type.  
High mass He star remnants have relatively shorter lives, 
so the numbers of such systems are predicted to decline with higher Be star mass. 
On the other hand, lower mass remnants above the He-burning limit ($\approx 0.3 M_\odot$)
will have He-burning lifetimes longer than those of the rejuvenated gainer stars, so 
many lower mass Be stars could have He star companions.  However, the expected numbers
depend critically on assumptions about the initial mass ratio distribution of the 
binaries and which systems merge during interaction.  \citet{pols1991} show in 
their Figure~10(b) the numbers of Be+He systems expected for a magnitude limited 
sample like the Bright Star Catalogue \citep{hoffleit1991}.  If the initial 
mass ratio distribution is flat, then the numbers of Be+He systems peak in the 
B0-B5 range, because there are relatively fewer low mass systems.  However, for 
initial mass ratio distributions that favor lower mass companions, the relative 
numbers of low mass Be+He systems increase.  This is also seen in the simulations 
by \citet{shao2014} who find the largest numbers of Be+He systems (in a volume limited sample)
among the latest B-types when the assumed initial mass ratio distribution
is highly skewed to low mass companions (numbers proportional to $(M_2/M_1)^{-1}$; see their
Fig.\ 7).  The observed lack of Be+sdO companions among the late-type B-stars in our sample 
would appear to favor progenitor systems with a flat initial mass ratio distribution.  

The low-mass, rapid rotators that comprise the late-type Be stars may be the result of mergers 
that occur more frequently among lower mass systems \citep{shao2014} or they may result from extreme 
mass transfer that creates low mass cores that quickly cool to become helium white dwarfs \citep{chen2017}.  
Such low mass white dwarfs are now known to orbit some late B-type, rapid rotators including Regulus 
($0.3 M_\odot$; \citealt{gies2008}), KOI-81 ($0.2 M_\odot$; \citealt{matson2015}), and
possibly $\beta$~CMi ($0.4 M_\odot$; \citealt{dulaney2017}).

If binary mass transfer is a significant mechanism in the production of Be stars, then 
we need to consider why hot companions are not found for all the B0-B3 stars in our sample. 
We suspect that many such companions spend much of their lives with a luminosity that 
is too low for detection by our methods.  For example, \citet{schootemeijer2017} present 
a path in the H-R diagram for the stripped remnant of HD~10516 ($\phi$~Per) 
based on He star evolutionary models.  They argue that the subdwarf companions 
in $\phi$~Per, FY~CMa, and 59~Cyg are most likely experiencing a helium shell burning stage, 
in which the core has finished helium burning and the star swells to increased luminosity. 
This phase lasts about 10$\%$ of the total post-mass transfer lifetime.  Thus, we expect that 
those stars that are bright enough to detect in our survey are representative of this advanced 
He-shell burning stage.  The fraction of detected companions in B0-B3 range ($17/140 = 12\%$) 
is close to the typical fraction of time spent in He-shell burning phase.  On the other hand, 
90$\%$ of the post-mass transfer lifetime is spent in the prior He-core burning stage, during which 
the subdwarfs have lower luminosity.  These faint sdO companions are very difficult to detect unless 
a large set of spectra is available to reveal the orbital motion (for example, the case of 
HR~2142; \citealt{peters2016}).  Consequently, it is possible that many or most of the B0-B3 
stars in our sample do have sdO companions that are undetected because they are faint, He-core 
burning objects.  Long term radial velocity and orbital-phase related emission line monitoring 
\citep{koubsky2012} may prove to be effective ways to discover their binary properties.

%%%%%%%%%%%%%%%%%%%%%%%%%%%%%%%%%%%%%%%%%%%%%%%%%%%%%%%%%%%%%%%%%%%%%%%%%%%%

\acknowledgments

We thank Dr.\ Michael Crenshaw for helping us interpret the radial velocity shifts apparent 
in some of the \emph{IUE} spectra, and we thank an anonymous referee for their insight 
about the spectral subtype distribution of the sample. 
The data presented in this paper were obtained from the Mikulski Archive for Space Telescopes (MAST). 
STScI is operated by the Association of Universities for Research in Astronomy, Inc., under 
NASA contract NAS5-26555. Support for MAST for non-HST data is provided by the 
NASA Office of Space Science via grant NNX09AF08G and by other grants and contracts. 
Our work was supported in part by NASA grant NNX10AD60G (GJP) and by the 
National Science Foundation under grant AST-1411654 (DRG). 
Institutional support has been provided from the GSU College of Arts and Sciences, 
the Research Program Enhancement fund of the Board of Regents of the University System of Georgia 
(administered through the GSU Office of the Vice President for Research and Economic Development), 
and by the USC Women in Science and Engineering (WiSE) program (GJP). 

%%%%%%%%%%%%%%%%%%%%%%%%%%%%%%%%%%%%%%%%%%%%%%%%%%%%%%%%%%%%%%%

Facilities: \facility{IUE}

%%%%%%%%%%%%%%%%%%%%%%%%%%%%%%%%%%%%%%%%%%%%%%%%%%%%%%%%%%%%%%%

\newpage

\appendix
\section{Notes on individual stars}
% HD 35411
\noindent {\bf{HD 35411 ($\eta$ Ori; B1 \rom{4})}}. \citet{zizka1981} measured 
the spectroscopic radial velocities of the inner binary system and concluded that 
the Aa1, Aa2 pair has an orbital period of 7.989~d. The first speckle measurements 
done by \citet{mcalister1976} showed that there is a third star (B3 \rom{4}) that 
orbits around the inner eclipsing and spectroscopic binary system 
Aa1, Aa2 (B1 \rom{4} and B3 \rom{4}, respectively) in an orbit with an angular 
semimajor axis of 44 mas and a period of 9.2 yr. The \emph{IUE} spectra recorded the 
flux of all three components. The cross-correlation functions of the spectra show 
a sharp peak, and the peak half-width is comparable to the projected rotational 
velocity of Aa1. Furthermore, the measured CCF peak velocities appear to match the 
Aa1 radial velocity curve from \citet{demey1996} in their Figure~3. Thus, we conclude 
that the CCF peaks result from correlation with the spectral features of the 
primary Aa1 component of the inner binary system.  \newline
% HD 53367
{\bf{HD 53367 (V750 Mon; B0 \rom{4}e)}}. Based on spectroscopic studies of the 
optical lines, \citet{pogodin2006} proposed that the system consists of a primary 
main-sequence star ($\sim20 M_{\odot}$) and a pre-main sequence secondary ($\sim5 M_{\odot}$). 
We compared the measured peak velocities with the primary radial velocity curve in 
Figure~6 of \citet{pogodin2006}, and the similarity of the velocities suggests that 
the CCF peak is mostly due to correlation with the spectral features of the primary component. 
However, we noticed that two spectra (SWP38685 and SWP38686) yielded a peak velocity variation 
of $\sim 50$ km~s$^{-1}$ in less than 3 hours. We examined the spectra and found that the star 
drifted in position across the dispersion in the large aperture of the camera between exposures. 
Thus, this rapid velocity variation is instrumental in origin. \newline
% HD 135160  
{\bf{HD 135160 (B0 \rom{4})}}. The CCFs show both a sharp peak and an extended wing feature 
that varies in velocity. \citet{chini2012} detected spectral lines of both components of this 
suspected spectroscopic binary. The Doppler shifts apparent in the CCFs tend to confirm the 
spectroscopic binary nature of the system. \newline
% HD 166596
{\bf{HD 166596 (V692 CrA; B2.5 \rom{3}p)}}. {\emph{IUE}} recorded two spectra of this star 
within $\sim$1 hour. The peak velocities of the CCFs indicate a significant shift of 
$\sim 37$ km~s$^{-1}$ between the observations. \citet{renson2009} report that the target 
is a silicon star with a rotational period of $\sim 1.7$~d. We suggest that the rapid velocity 
variation is probably related to rotational Doppler shifts of regions with chemical peculiarities. \newline
% HD 178175
{\bf{HD 178175 (V4024 Sgr; B2 \rom{4}(e))}}. The CCFs show a sharp peak. Based on 
spectroscopic studies, \citet{braganca2012} reported that the star has $V\sin{i} = 86$ km s$^{-1}$, 
similar to the half-width of the CCF peak. However, \citet{braganca2012} estimate that the star's 
temperature is $T_{\rm eff} = 19.6$ kK, and we would usually expect little correlation with 
the features of such a relatively cooler object. However, the CCFs show no significant peak 
velocity variations. Thus, we conclude that the CCFs are most likely due to correlation with 
the features of the Be star that are narrow enough in this case to produce a detectable CCF peak. \newline
 
\newpage

%%%%%%%%%%%%%%%%%%%%%%%%%%%%%%%%%%%%%%%%%%%%%%%%%%%%%%%%%%%%%%%%
% Bibliography

\bibliographystyle{apj}
\bibliography{apj-jour,ms}

\begin{thebibliography}{64}
\expandafter\ifx\csname natexlab\endcsname\relax\def\natexlab#1{#1}\fi

\bibitem[{{Abt} {et~al.}(2002){Abt}, {Levato}, \& {Grosso}}]{abt2002}
{Abt}, H.~A., {Levato}, H., \& {Grosso}, M. 2002, \apj, 573, 359

\bibitem[{{Abt} \& {Levy}(1978)}]{abt1978}
{Abt}, H.~A., \& {Levy}, S.~G. 1978, \apjs, 36, 241

\bibitem[{{Althaus} {et~al.}(2013){Althaus}, {Miller Bertolami}, \&
  {C{\'o}rsico}}]{althaus2013}
{Althaus}, L.~G., {Miller Bertolami}, M.~M., \& {C{\'o}rsico}, A.~H. 2013,
  \aap, 557, A19

\bibitem[{{Baade} {et~al.}(2017){Baade}, {Pigulski}, {Rivinius}, {Carciofi},
  {Panoglou}, {Ghoreyshi}, {Handler}, {Kuschnig}, {Moffat}, {Pablo},
  {Popowicz}, {Wade}, {Weiss}, \& {Zwintz}}]{baade2017}
{Baade}, D., {Pigulski}, A., {Rivinius}, T., {et~al.} 2017, \aap, submitted
  (arXiv 1708.07360)

\bibitem[{{Bragan{\c c}a} {et~al.}(2012){Bragan{\c c}a}, {Daflon}, {Cunha},
  {Bensby}, {Oey}, \& {Walth}}]{braganca2012}
{Bragan{\c c}a}, G.~A., {Daflon}, S., {Cunha}, K., {et~al.} 2012, \aj, 144, 130

\bibitem[{{Bruzual} \& {Charlot}(2003)}]{bruzual2003}
{Bruzual}, G., \& {Charlot}, S. 2003, \mnras, 344, 1000

\bibitem[{{Cantiello} {et~al.}(2007){Cantiello}, {Yoon}, {Langer}, \&
  {Livio}}]{cantiello2007}
{Cantiello}, M., {Yoon}, S.-C., {Langer}, N., \& {Livio}, M. 2007, \aap, 465,
  L29

\bibitem[{{Chen} {et~al.}(2017){Chen}, {Maxted}, {Li}, \& {Han}}]{chen2017}
{Chen}, X., {Maxted}, P.~F.~L., {Li}, J., \& {Han}, Z. 2017, \mnras, 467, 1874

\bibitem[{{Chini} {et~al.}(2012){Chini}, {Hoffmeister}, {Nasseri}, {Stahl}, \&
  {Zinnecker}}]{chini2012}
{Chini}, R., {Hoffmeister}, V.~H., {Nasseri}, A., {Stahl}, O., \& {Zinnecker},
  H. 2012, \mnras, 424, 1925

\bibitem[{{Chojnowski} {et~al.}(2017){Chojnowski}, {Wisniewski}, {Whelan},
  {Labadie-Bartz}, {Borges Fernandes}, {Lin}, {Majewski}, {Stringfellow},
  {Mennickent}, {Roman-Lopes}, {Tang}, {Hearty}, {Holtzman}, {Pepper}, \&
  {Zasowski}}]{chojnowski2017}
{Chojnowski}, S.~D., {Wisniewski}, J.~P., {Whelan}, D.~G., {et~al.} 2017, \aj,
  153, 174

\bibitem[{{de Bruijne} \& {Eilers}(2012)}]{bruijne2012}
{de Bruijne}, J.~H.~J., \& {Eilers}, A.-C. 2012, \aap, 546, A61

\bibitem[{{De Mey} {et~al.}(1996){De Mey}, {Aerts}, {Waelkens}, \& {Van
  Winckel}}]{demey1996}
{De Mey}, K., {Aerts}, C., {Waelkens}, C., \& {Van Winckel}, H. 1996, \aap,
  310, 164

\bibitem[{{de Mink} {et~al.}(2014){de Mink}, {Sana}, {Langer}, {Izzard}, \&
  {Schneider}}]{demink2014}
{de Mink}, S.~E., {Sana}, H., {Langer}, N., {Izzard}, R.~G., \& {Schneider},
  F.~R.~N. 2014, \apj, 782, 7

\bibitem[{{Dulaney} {et~al.}(2017){Dulaney}, {Richardson}, {Gerhartz},
  {Bjorkman}, {Bjorkman}, {Carciofi}, {Klement}, {Wang}, {Morrison},
  {Bratcher}, {Greco}, {Hardegree-Ullman}, {Lembryk}, {Oswald}, \&
  {Trucks}}]{dulaney2017}
{Dulaney}, N.~A., {Richardson}, N.~D., {Gerhartz}, C.~J., {et~al.} 2017, \apj,
  836, 112

\bibitem[{{Eldridge} {et~al.}(2013){Eldridge}, {Fraser}, {Smartt}, {Maund}, \&
  {Crockett}}]{eldridge2013}
{Eldridge}, J.~J., {Fraser}, M., {Smartt}, S.~J., {Maund}, J.~R., \&
  {Crockett}, R.~M. 2013, \mnras, 436, 774

\bibitem[{{Fr{\'e}mat} {et~al.}(2005){Fr{\'e}mat}, {Zorec}, {Hubert}, \&
  {Floquet}}]{fremat2005}
{Fr{\'e}mat}, Y., {Zorec}, J., {Hubert}, A.-M., \& {Floquet}, M. 2005, \aap,
  440, 305

\bibitem[{{Gies} {et~al.}(1998){Gies}, {Bagnuolo}, {Ferrara}, {Kaye},
  {Thaller}, {Penny}, \& {Peters}}]{gies1998}
{Gies}, D.~R., {Bagnuolo}, Jr., W.~G., {Ferrara}, E.~C., {et~al.} 1998, \apj,
  493, 440

\bibitem[{{Gies} {et~al.}(2008){Gies}, {Dieterich}, {Richardson}, {Riedel},
  {B.~L.~Team}, {McAlister}, {Bagnuolo}, {Grundstrom}, {{\v S}tefl},
  {Rivinius}, \& {Baade}}]{gies2008}
{Gies}, D.~R., {Dieterich}, S., {Richardson}, N.~D., {et~al.} 2008, \apjl, 682,
  L117

\bibitem[{{Gim{\'e}nez-Garc{\'{\i}}a}
  {et~al.}(2015){Gim{\'e}nez-Garc{\'{\i}}a}, {Torrej{\'o}n}, {Eikmann},
  {Mart{\'{\i}}nez-N{\'u}{\~n}ez}, {Oskinova}, {Rodes-Roca}, \&
  {Bernab{\'e}u}}]{garcia2015}
{Gim{\'e}nez-Garc{\'{\i}}a}, A., {Torrej{\'o}n}, J.~M., {Eikmann}, W., {et~al.}
  2015, \aap, 576, A108

\bibitem[{{Han} {et~al.}(2010){Han}, {Podsiadlowski}, \&
  {Lynas-Gray}}]{han2010}
{Han}, Z., {Podsiadlowski}, P., \& {Lynas-Gray}, A. 2010, \apss, 329, 41

\bibitem[{{Hartkopf} {et~al.}(1996){Hartkopf}, {Mason}, {McAlister}, {Turner},
  {Barry}, {Franz}, \& {Prieto}}]{hartkopf1996}
{Hartkopf}, W.~I., {Mason}, B.~D., {McAlister}, H.~A., {et~al.} 1996, \aj, 111,
  936

\bibitem[{{Hoffleit} \& {Jaschek}(1991)}]{hoffleit1991}
{Hoffleit}, D., \& {Jaschek}, C. 1991, {The Bright Star Catalogue} (5th ed.;
  New Haven, CT: Yale University Observatory)

\bibitem[{{Hoogerwerf} {et~al.}(2001){Hoogerwerf}, {de Bruijne}, \& {de
  Zeeuw}}]{hoogerwerf2001}
{Hoogerwerf}, R., {de Bruijne}, J.~H.~J., \& {de Zeeuw}, P.~T. 2001, \aap, 365,
  49

\bibitem[{{Houk}(1978)}]{houk1978}
{Houk}, N. 1978, {Michigan Catalogue of Two-dimensional Spectral Types for the
  HD stars. Declinations $-52$ to $-41$}, Vol.~2 (Ann Arbor, MI: University of
  Michigan)

\bibitem[{{Houk}(1982)}]{houk1982}
---. 1982, {Michigan Catalogue of Two-dimensional Spectral Types for the HD
  stars. Declinations $-40$ to $-26$}, Vol.~3 (Ann Arbor, MI: University of
  Michigan)

\bibitem[{{Houk} \& {Cowley}(1975)}]{houk1975}
{Houk}, N., \& {Cowley}, A.~P. 1975, {Michigan Catalogue of Two-dimensional
  Spectral Types for the HD stars. Declinations $-90$ to $-53$}, Vol.~1 (Ann
  Arbor, MI: University of Michigan)

\bibitem[{{Houk} \& {Swift}(1999)}]{houk1999}
{Houk}, N., \& {Swift}, C. 1999, {Michigan Catalogue of Two-dimensional
  Spectral Types for the HD stars. Declinations $-12$ to $5$}, Vol.~5 (Ann
  Arbor, MI: University of Michigan)

\bibitem[{{Huang} \& {Gies}(2006)}]{huang2006}
{Huang}, W., \& {Gies}, D.~R. 2006, \apj, 648, 580

\bibitem[{{Huang} {et~al.}(2010){Huang}, {Gies}, \& {McSwain}}]{huang2010}
{Huang}, W., {Gies}, D.~R., \& {McSwain}, M.~V. 2010, \apj, 722, 605

\bibitem[{{Koubsk{\'y}} {et~al.}(2012){Koubsk{\'y}}, {Kotkov{\'a}}, {Votruba},
  {{\v S}lechta}, \& {Dvo{\v r}{\'a}kov{\'a}}}]{koubsky2012}
{Koubsk{\'y}}, P., {Kotkov{\'a}}, L., {Votruba}, V., {{\v S}lechta}, M., \&
  {Dvo{\v r}{\'a}kov{\'a}}, {\v S}. 2012, \aap, 545, A121

\bibitem[{{Koubsk{\'y}} {et~al.}(2000){Koubsk{\'y}}, {Harmanec}, {Hubert},
  {Floquet}, {Kub{\'a}t}, {Ballereau}, {Chauville}, {Bo{\v z}i{\'c}},
  {Holmgren}, {Yang}, {Cao}, {Eenens}, {Huang}, \& {Percy}}]{koubsky2000}
{Koubsk{\'y}}, P., {Harmanec}, P., {Hubert}, A.~M., {et~al.} 2000, \aap, 356,
  913

\bibitem[{{Lanz} \& {Hubeny}(2003)}]{lanz2003}
{Lanz}, T., \& {Hubeny}, I. 2003, \apjs, 146, 417

\bibitem[{{Lesh}(1968)}]{lesh1968}
{Lesh}, J.~R. 1968, \apjs, 17, 371

\bibitem[{{Levenhagen} \& {Leister}(2006)}]{lavenhagen2006}
{Levenhagen}, R.~S., \& {Leister}, N.~V. 2006, \mnras, 371, 252

\bibitem[{{Lindroos}(1985)}]{lindroos1985}
{Lindroos}, K.~P. 1985, \aaps, 60, 183

\bibitem[{{Lopes de Oliveira} \& {Motch}(2011)}]{oliveira2011}
{Lopes de Oliveira}, R., \& {Motch}, C. 2011, \apjl, 731, L6

\bibitem[{{Mason} {et~al.}(2007){Mason}, {Hartkopf}, {Wycoff}, \&
  {Wieder}}]{mason2007}
{Mason}, B.~D., {Hartkopf}, W.~I., {Wycoff}, G.~L., \& {Wieder}, G. 2007, \aj,
  134, 1671

\bibitem[{{Mason} {et~al.}(1997){Mason}, {ten Brummelaar}, {Gies}, {Hartkopf},
  \& {Thaller}}]{mason1997}
{Mason}, B.~D., {ten Brummelaar}, T., {Gies}, D.~R., {Hartkopf}, W.~I., \&
  {Thaller}, M.~L. 1997, \aj, 114, 2112

\bibitem[{{Mason} {et~al.}(2001){Mason}, {Wycoff}, {Hartkopf}, {Douglass}, \&
  {Worley}}]{mason2001}
{Mason}, B.~D., {Wycoff}, G.~L., {Hartkopf}, W.~I., {Douglass}, G.~G., \&
  {Worley}, C.~E. 2001, \aj, 122, 3466

\bibitem[{{Matson} {et~al.}(2015){Matson}, {Gies}, {Guo}, {Quinn}, {Buchhave},
  {Latham}, {Howell}, \& {Rowe}}]{matson2015}
{Matson}, R.~A., {Gies}, D.~R., {Guo}, Z., {et~al.} 2015, \apj, 806, 155

\bibitem[{{Mayer} {et~al.}(2016){Mayer}, {Deschamps}, \&
  {Jorissen}}]{mayer2016}
{Mayer}, A., {Deschamps}, R., \& {Jorissen}, A. 2016, \aap, 587, A30

\bibitem[{{McAlister}(1976)}]{mcalister1976}
{McAlister}, H.~A. 1976, \pasp, 88, 957

\bibitem[{{Peters} {et~al.}(2008){Peters}, {Gies}, {Grundstrom}, \&
  {McSwain}}]{peters2008}
{Peters}, G.~J., {Gies}, D.~R., {Grundstrom}, E.~D., \& {McSwain}, M.~V. 2008,
  \apj, 686, 1280

\bibitem[{{Peters} {et~al.}(2013){Peters}, {Pewett}, {Gies}, {Touhami}, \&
  {Grundstrom}}]{peters2013}
{Peters}, G.~J., {Pewett}, T.~D., {Gies}, D.~R., {Touhami}, Y.~N., \&
  {Grundstrom}, E.~D. 2013, \apj, 765, 2

\bibitem[{{Peters} {et~al.}(2016){Peters}, {Wang}, {Gies}, \&
  {Grundstrom}}]{peters2016}
{Peters}, G.~J., {Wang}, L., {Gies}, D.~R., \& {Grundstrom}, E.~D. 2016, \apj,
  828, 47

\bibitem[{{Plaskett} {et~al.}(1920){Plaskett}, {Harper}, {Young}, \&
  {Plaskett}}]{plaskett1920}
{Plaskett}, J.~S., {Harper}, W.~E., {Young}, R.~K., \& {Plaskett}, H.~H. 1920,
  Publications of the Dominion Astrophysical Observatory Victoria, 1, 163

\bibitem[{{Pogodin} {et~al.}(2012){Pogodin}, {Drake}, {Jilinski}, {Daflon},
  {Herencia}, {de la Reza}, \& {Ortega}}]{pogodin2012}
{Pogodin}, M.~A., {Drake}, N.~A., {Jilinski}, E.~G., {et~al.} 2012, in
  Circumstellar Dynamics at High Resolution, ASP Conf. Vol. 464, ed. A.~C.
  {Carciofi} \& T.~{Rivinius} (San Francisco, CA: ASP), 227

\bibitem[{{Pogodin} {et~al.}(2006){Pogodin}, {Malanushenko}, {Kozlova},
  {Tarasova}, \& {Franco}}]{pogodin2006}
{Pogodin}, M.~A., {Malanushenko}, V.~P., {Kozlova}, O.~V., {Tarasova}, T.~N.,
  \& {Franco}, G.~A.~P. 2006, \aap, 452, 551

\bibitem[{{Pols} {et~al.}(1991){Pols}, {Cote}, {Waters}, \& {Heise}}]{pols1991}
{Pols}, O.~R., {Cote}, J., {Waters}, L.~B.~F.~M., \& {Heise}, J. 1991, \aap,
  241, 419

\bibitem[{{Renson} \& {Manfroid}(2009)}]{renson2009}
{Renson}, P., \& {Manfroid}, J. 2009, \aap, 498, 961

\bibitem[{{Rivinius} {et~al.}(2013){Rivinius}, {Carciofi}, \&
  {Martayan}}]{rivinius2013}
{Rivinius}, T., {Carciofi}, A.~C., \& {Martayan}, C. 2013, \aapr, 21, 69

\bibitem[{{Sana} {et~al.}(2012){Sana}, {de Mink}, {de Koter}, {Langer},
  {Evans}, {Gieles}, {Gosset}, {Izzard}, {Le Bouquin}, \&
  {Schneider}}]{sana2012}
{Sana}, H., {de Mink}, S.~E., {de Koter}, A., {et~al.} 2012, Science, 337, 444

\bibitem[{{Schootemeijer} {et~al.}(2017){Schootemeijer}, {G{\" o}tberg}, {de
  Mink}, {Gies}, \& {Zapartas}}]{schootemeijer2017}
{Schootemeijer}, A., {G{\" o}tberg}, Y., {de Mink}, S.~E., {Gies}, D., \&
  {Zapartas}, E. 2017, \aap, submitted

\bibitem[{{Shao} \& {Li}(2014)}]{shao2014}
{Shao}, Y., \& {Li}, X.-D. 2014, \apj, 796, 37

\bibitem[{{Shatskii}(1998)}]{shatskii1998}
{Shatskii}, N.~I. 1998, Astronomy Letters, 24, 257

\bibitem[{{Sim{\'o}n-D{\'{\i}}az} \& {Herrero}(2014{\natexlab{a}})}]{simon2014}
{Sim{\'o}n-D{\'{\i}}az}, S., \& {Herrero}, A. 2014{\natexlab{a}}, \aap, 562,
  A135

\bibitem[{{Sim{\'o}n-D{\'{\i}}az} \&
  {Herrero}(2014{\natexlab{b}})}]{simondiaz2014}
---. 2014{\natexlab{b}}, \aap, 562, A135

\bibitem[{{Slettebak}(1966)}]{slettebak1966}
{Slettebak}, A. 1966, \apj, 145, 126

\bibitem[{{Slettebak}(1982)}]{slettebak1982}
---. 1982, \apjs, 50, 55

\bibitem[{{Thaller} {et~al.}(1995){Thaller}, {Bagnuolo}, {Gies}, \&
  {Penny}}]{thaller1995}
{Thaller}, M.~L., {Bagnuolo}, Jr., W.~G., {Gies}, D.~R., \& {Penny}, L.~R.
  1995, \apj, 448, 878

\bibitem[{{Uesugi} \& {Fukuda}(1970)}]{uesugi1970}
{Uesugi}, A., \& {Fukuda}, I. 1970, {Catalogue of rotational velocities of the
  stars} (Kyoto: University of Kyoto)

\bibitem[{{Wang} {et~al.}(2017){Wang}, {Gies}, \& {Peters}}]{wang2017}
{Wang}, L., {Gies}, D.~R., \& {Peters}, G.~J. 2017, \apj, 843, 60

\bibitem[{{Yudin}(2001)}]{yudin2001}
{Yudin}, R.~V. 2001, \aap, 368, 912

\bibitem[{{Zizka} \& {Beardsley}(1981)}]{zizka1981}
{Zizka}, E.~R., \& {Beardsley}, W.~R. 1981, \aj, 86, 1944

\end{thebibliography}
%\bibliography{ms}

%%%%%%%%%%%%%%%%%%%%%%%%%%%%%%%%%%%%%%%%%%%%%%%%%%%%%%%%%%%%%%

% Table 1: List of low S/N stars
\begin{deluxetable}{ccccc}
\tabletypesize{\scriptsize}
\tablenum{1}
\tablecaption{Stars with CCF S/N $<$ 3}
\tablewidth{0pt}
\tablehead{
\colhead{\phn Name} & 
\colhead{\phn Name} & 
\colhead{\phn Name} &
\colhead{\phn Name} & 
\colhead{Name}}
\startdata
\phn\phn\phn144& \phn\phn4180 & \phn\phn6811 & \phn10144 & \phn11415   \\
\phn11946& \phn13268& \phn14434& \phn15642& \phn18552  \\
\phn20340& \phn21362&\phn21551& \phn22192& \phn22780  \\
\phn23016& \phn23302&\phn23383&\phn23478& \phn23480 \\
\phn23552& \phn23630&\phn23862&\phn24479&\phn25940  \\
\phn26356&\phn26670&\phn26793&\phn28459&\phn28867 \\
\phn29866&\phn32343&\phn32990&\phn32991&\phn33599 \\
\phn34863&\phn34959&\phn35407&\phn35439&\phn36012 \\
\phn36408&\phn36665&\phn36939&\phn37115&\phn37202 \\
\phn37397&\phn37795&\phn37967&\phn38087&\phn38831 \\
{\phn\phn41335}$^{*}$&\phn42054&\phn42545&\phn44458&\phn44506 \\
\phn45314&\phn45542&\phn45725&\phn45910&\phn45995 \\
\phn46056&\phn46485&\phn47054&\phn47359&\phn50123 \\
\phn50820&\phn51480&\phn52356&\phn52721&\phn56139 \\
\phn57150&\phn58343&\phn58715&\phn60606&\phn61355 \\
\phn63462&\phn65875&\phn69106&\phn69404&\phn70084 \\
\phn71216&\phn72014&\phn72067&\phn75311&\phn75416 \\
\phn76534&\phn77366&\phn79351&\phn83953&\phn86612 \\
\phn87543&\phn87901&\phn89080&\phn89884&\phn89890 \\
\phn91120&\phn91465&\phn92938&\phn93563&100673 \\
100889&102776&105382&107348&109387 \\
109857&110335&110432&110863&112078 \\
112091&118246&119921&121847&124367 \\
127973&130109&134481&135734&137432 \\
138749&138769&139431&141637&142184 \\
142926&142983&149485&149671&149757 \\
155896&156468&158427&158643&160202 \\
162732&164284&164906&165063&166014 \\
167128&168957&169033&171406&174639 \\
175869&177724&178475&179343&181615 \\
182180&183133&183362&183656&183914 \\
185037&187235&187811&189687&191423 \\
192044&192685&192954&193182&193911 \\
195325&198183&198625&199218&202904 \\
203064&203467&203699&204860&205637 \\
206773&208057&208392&208682&208886 \\
209014&209409&209522&210129&214748 \\
216200&217050&217086&217543&217676 \\
217891&218393&218674&219688&224686 \\
BD+41 3731 &  BD+60 594 &  BD+34 1058 &  &  
\enddata
\tablenotetext{*}{HR~2142 has a faint sdO companion \citep{peters2016}.}
\tablecomments{Stars are listed by HD number, except for the last three.}
\end{deluxetable}

% Table 2: Summary of the targets' observations
\begin{deluxetable}{cccccccccc}
\rotate
\tabletypesize{\scriptsize}
\tablenum{2}
\tablecaption{\emph{IUE} Observations of sample stars}
\tablewidth{0pt}
\tablehead{
\colhead{HD} & 
\colhead{Star} & 
\colhead{HIP} & 
\colhead{Spectral} & 
\colhead{$V\sin i$} &      
\colhead{Number of}  & 
\colhead{S/N} &
\colhead{Spectral classification} & 
\colhead{$V\sin i$} &
\colhead{\phn CCF} \\
\colhead{Number} & 
\colhead{Name} & 
\colhead{Number} & 
\colhead{Classification} & 
\colhead{(km s$^{-1}$)} &
\colhead{Observations} &
\colhead{Ratio}  &
\colhead{Reference} &
\colhead{Reference} & 
\colhead{Code$^{a}$} }
\startdata
\multicolumn{10}{c}{Candidate and Known Be+sdO Systems} \\
\hline
\phn10516    & $\phi$ Per   & \phn\phn8068  & B1.5 \rom{5}:e-shell  & 440     & \phn16    & 6.78      & \citet{slettebak1982} & \citet{fremat2005}   & S  \\
\phn29441    & V1150 Tau    & \phn21626     & B2.5 \rom{5}ne        & 311     & \phn\phn1 & 3.77      & \citet{yudin2001}     & \citet{yudin2001}    & C? \\
\phn43544    & \nodata      & \phn29771     & B3 \rom{5}            & 256     & \phn\phn1 & 5.51      & \citet{slettebak1982} & \citet{braganca2012} & C? \\ 
\phn51354    & QY Gem       & \phn33493     & B3 \rom{5}e           & 306     & \phn\phn2 & 5.39      & \citet{slettebak1982} & \citet{yudin2001}    & C? \\    
\phn58978    & FY CMa       & \phn36168     & B0.5 \rom{4}e         & 340     & \phn96    & 3.10      & \citet{slettebak1982} & \citet{peters2008}   & S  \\
\phn60855    & V378 Pup     & \phn36981     & B3 \rom{4}            & 239     & \phn\phn6 & 4.98      & \citet{slettebak1982} & \citet{fremat2005}   & C  \\
113120       & LS Mus       & \phn63688     & B2 \rom{4}ne          & 307     & \phn\phn3 & 4.35      & \citet{lavenhagen2006}& \citet{yudin2001}    & C  \\ 
137387       & $\kappa^{1}$ Aps&\phn76013   & B2 \rom{5}npe         & 250     & \phn\phn4 & 5.21      & \citet{lavenhagen2006}& \citet{fremat2005}   & C  \\
152478       & V846 Ara     & \phn82868     & B3 \rom{5}npe         & 340     & \phn\phn2 & 3.69      & \citet{lavenhagen2006}& \citet{pogodin2012}  & C  \\  
157042       & $\iota$ Ara  & \phn85079     & B2.5 \rom{4}e         & 340     & \phn\phn4 & 4.09      & \citet{slettebak1982} & \citet{fremat2005}   & C  \\
157832       & V750 Ara     & \phn85467     & B1.5 \rom{5}e         & 266     & \phn\phn2 & 4.21      & \citet{oliveira2011}  & \citet{oliveira2011} & C? \\
191610       & 28 Cyg       & \phn99303     & B3 \rom{4}e           & 300     & \phn46    & 3.02      & \citet{slettebak1982} & \citet{fremat2005}   & C  \\
194335       & V2119 Cyg    & 100574        & B2 \rom{3}e           & 360     & \phn\phn4 & 4.54      & \citet{slettebak1982} & \citet{fremat2005}   & C  \\
200120       & 59 Cyg       & 103632        & B1 \rom{5}e           & 379     &  193      & 3.34      & \citet{slettebak1982} & \citet{fremat2005}   & S  \\
200310       & 60 Cyg       & 103732        & B1 \rom{5}e           & 320     & \phn23    & 4.69      & \citet{koubsky2000}   & \citet{koubsky2000}  & S  \\
214168       & 8 Lac B      & 111544        & B1 \rom{4}e           & 300     & \phn20    & 3.03      & \citet{slettebak1982} & \citet{fremat2005}   & C  \\
\hline
\multicolumn{10}{c}{Other Stars} \\
\hline
\phn\phn5394 & $\gamma$ Cas & \phn\phn4427  & B0.5 \rom{4}e         & 432     &  227      & 3.81      & \citet{slettebak1982} & \citet{fremat2005}   & P  \\
\phn20336    & BK Cam       & \phn15520     & B2 (\rom{4}:)e  	    & 328     & \phn\phn2 & 3.31      & \citet{slettebak1982} & \citet{fremat2005}   & P  \\
\phn24534    & X Per        & \phn18350     & O9.5 \rom{3}          & 293     & \phn43    & 3.75      & \citet{slettebak1982} & \citet{fremat2005}   & P  \\
\phn28497    & DU Eri       & \phn20922     & B1 \rom{5}e           & 300     & \phn28    & 3.01      & \citet{slettebak1982} & \citet{fremat2005}   & P  \\
\phn30076    & DX Eri       & \phn22024     & B2 \rom{5}e           & 168     & \phn24    & 3.30      & \citet{slettebak1982} & \citet{huang2010}    & P  \\
\phn33328    & $\lambda$ Eri& \phn23972     & B2 \rom{3}(e)p        & 318     & 146       & 3.35      & \citet{slettebak1982} & \citet{fremat2005}   & P  \\
\phn35411    & $\eta$ Ori   & \phn25281     & B1 \rom{5}            &\phn20   & \phn19    & 10.28\phn & \citet{slettebak1982} & \citet{demey1996}    & P  \\ 
\phn36576    & V960 Tau     & \phn26064     & B1.5 \rom{4}e         & 265     & \phn11    & 3.14      & \citet{slettebak1982} & \citet{fremat2005}   & P  \\
\phn37490    & $\omega$ Ori & \phn26594     & B3 \rom{5}e           & 171     &  190      & 3.43      & \citet{lavenhagen2006}& \citet{yudin2001}    & P  \\ 
\phn37674    & \nodata	    & \phn26683     & B5 \rom{5}(n)         & \nodata & \phn\phn1 & 3.34      & \citet{houk1999}      & \nodata              & P  \\
\phn48917    & FT CMa       & \phn32292     & B2 \rom{5}            & 205     & \phn\phn5 & 3.48      & \citet{houk1982}      & \citet{fremat2005}   & P  \\
\phn50013    & $\kappa$ CMa & \phn32759     & B2 \rom{4}e           & 243     & \phn\phn3 & 4.55      & \citet{slettebak1982} & \citet{fremat2005}   & P  \\
\phn50083    & V742 Mon     & \phn32947     & B2 \rom{5}e           & 170     & \phn\phn2 & 3.90      & \citet{yudin2001}     & \citet{fremat2005}   & P  \\
\phn52918    & 19 Mon       & \phn33971     & B2 \rom{5}n(e)        & 265     & \phn15    & 3.10      & \citet{houk1999}      & \citet{huang2010}    & P  \\
\phn53367    & V750 Mon     & \phn34116     & B0 \rom{4}e           & \phn86  & \phn\phn4 & 11.05\phn & \citet{pogodin2006}   & \citet{yudin2001}    & P  \\
\phn54309    & FV CMa       & \phn34360     & B2 \rom{4}e           & 195     & \phn\phn2 & 3.63      & \citet{slettebak1982} & \citet{fremat2005}   & P  \\ 
\phn56014    & EW CMa       & \phn34981     & B3 \rom{3}(e)p-shell  & 280     & \phn12    & 3.45      & \citet{slettebak1982} & \citet{fremat2005}   & P  \\ 
\phn58050    & OT Gem       & \phn35933     & B2 \rom{5}e	    & 130     & \phn\phn4 & 3.65      & \citet{yudin2001}     & \citet{fremat2005}   & P  \\
\phn60848    & BN Gem       & \phn37074     & O8 \rom{5}pe          & 247     & \phn13    & 4.05      & \citet{yudin2001}     & \citet{fremat2005}   & P  \\  
\phn66194    & V374 Car     & \phn38994     & B2.5 \rom{4}e         & 224     & \phn\phn2 & 3.27      & \citet{slettebak1982} & \citet{yudin2001}    & P  \\
\phn67536    & V375 Car     & \phn39530     & B2 \rom{5}n           & 292     & \phn22    & 3.28      & \citet{houk1975}      & \citet{uesugi1970}   & P  \\
\phn68980    & MX Pup       & \phn40274     & B1.5 \rom{4}e         & 145     & \phn\phn3 & 4.91      & \citet{slettebak1982} & \citet{fremat2005}   & P  \\
\phn74455    & HX Vel       & \phn42712     & B2$/$3 \rom{4}$/$\rom{5}& 285   & \phn\phn7 & 4.56      & \citet{houk1978}      & \citet{uesugi1970}   & P  \\
\phn74753    & D Vel        & \phn42834     & B1$/$2 \rom{2}$/$\rom{3}(n)&288 & \phn\phn1 & 3.47      & \citet{houk1978}      & \citet{uesugi1970}   & P  \\
\phn78764    & E Car        & \phn44626     & B2 \rom{4}e           & 127     & \phn\phn3 & 5.42      & \citet{slettebak1982} & \citet{yudin2001}    & P  \\  
\phn88661    & QY Car       & \phn49934     & B2 \rom{4}e           & 237     & \phn14    & 3.71      & \citet{slettebak1982} & \citet{fremat2005}   & P  \\
\phn93030    & $\theta$ Car & \phn52419     & B0 \rom{5}p           & 145     & \phn33    & 8.68      & \citet{yudin2001}     & \citet{yudin2001}    & P  \\  
\phn96864    & \nodata      & \nodata       & B1.5 \rom{4}nep       & \nodata & \phn\phn1 & 4.16      & \citet{yudin2001}     & \nodata              & P  \\
105435	     & $\delta$ Cen & \phn59196     & B2 \rom{4}e           & 260     & \phn19    & 3.35      & \citet{slettebak1982} & \citet{fremat2005}   & P  \\  
116781	     & V967 Cen     & \phn65637     & B0 \rom{3}ne          & \nodata & \phn\phn1 & 7.55      & \citet{yudin2001}     & \nodata              & P  \\
120324       & $\mu$ Cen    & \phn67472     & B2 \rom{5}npe         & 159     & \phn36    & 4.62      & \citet{lavenhagen2006}& \citet{fremat2005}   & P  \\
120991       & V767 Cen     & \phn67861     & B2 \rom{3}ep          & \phn70  & \phn\phn5 & 6.09      & \citet{slettebak1982} & \citet{fremat2005}   & P  \\  
135160       & \nodata      & \phn74750     & B0 \rom{5}            & 155     & \phn\phn3 & 7.00      & \citet{slettebak1982} & \citet{yudin2001}    & P  \\  
148184       & $\chi$ Oph   & \phn80569     & B1.5 \rom{5}e         & 144     & \phn10    & 4.87      & \citet{slettebak1982} & \citet{fremat2005}   & P  \\ 
153261       & V828 Ara     & \phn83323     & B2 \rom{4}e           & 184     & \phn\phn1 & 3.92      & \citet{yudin2001}     & \citet{yudin2001}    & P  \\
155806       & V1075 Sco    & \phn84401     & O7.5 \rom{3}e         & 116     & \phn\phn6 & 9.96      & \citet{slettebak1982} & \citet{yudin2001}    & P  \\ 
166596       & V692 CrA     & \phn89290     & B2.5 \rom{3}p         & 207     & \phn\phn2 & 3.20      & \citet{yudin2001}     & \citet{fremat2005}   & P  \\
170235       & V4031 Sgr    & \phn90610     & B1 \rom{5}nne         & 163     & \phn\phn2 & 4.39      & \citet{lavenhagen2006}& \citet{yudin2001}    & P  \\ 
173219       & V447 Sct     & \phn91946     & B0 \rom{1}ae          & \nodata & \phn\phn3 & 4.15      & \citet{houk1999}      & \nodata              & P  \\
173948       & $\lambda$ Pav& \phn92609     & B2 \rom{5}e           & 140     & \phn\phn5 & 5.09      & \citet{lavenhagen2006}& \citet{fremat2005}   & P  \\
174237       & CX Dra       & \phn92133     & B4 \rom{4}(e)         & 163     & \phn64    & 3.16      & \citet{slettebak1982} & \citet{fremat2005}   & P  \\
178175       & V4024 Sgr    & \phn93996     & B2 \rom{5}(e)         & \phn86  & \phn\phn6 & 5.51      & \citet{slettebak1982} & \citet{braganca2012} & P  \\ 
184279       & V1294 Aql    & \phn96196     & B0 \rom{5}	    & 212     & \phn11    & 3.39      & \citet{lavenhagen2006}& \citet{yudin2001}    & P  \\   
184915       & $\kappa$ Aql & \phn96483     & B0.5 \rom{3}          & 284     & \phn\phn4 & 3.38      & \citet{slettebak1982} & \citet{simon2014}    & P  \\
187567       & V1339 Aql    & \phn97607     & B2.5 \rom{4}e         & 140     & \phn\phn3 & 3.26      & \citet{yudin2001}     & \citet{abt2002}      & P  \\ 
188439       & V819 Cyg     & \phn97845     & B0.5 \rom{3}(n)       & 299     & \phn\phn1 & 3.75      & \citet{lesh1968}      & \citet{simondiaz2014}& P  \\
203374       & \nodata      & 105250        & B0 \rom{4}pe          & 342     & \phn\phn1 & 3.76      & \citet{yudin2001}     & \citet{fremat2005}   & P  \\ 
212044       & V357 Lac     & 110287        & B0 \rom{5}e           & 162     & \phn\phn1 & 4.33      & \citet{yudin2001}     & \citet{yudin2001}    & P  \\ 
212076       & 31 Peg       & 110386        & B1.5 \rom{5}e         & \phn98  & \phn\phn6 & 3.95      & \citet{slettebak1982} & \citet{fremat2005}   & P  \\
212571       & $\pi$ Aqr    & 110672        & B1 \rom{5}e           & 230     & \phn21    & 3.18      & \citet{mayer2016}     & \citet{fremat2005}   & P  
\enddata   
%\tablecomments{Spectral classifications and projected rotational velocities are from 
%\citet{slettebak1982} and \citet{fremat2005}, respectively.}
\tablenotetext{a}{S = known Be+sdO binary; C = candidate binary; C? = potential candidate binary; P = CCF signal from primary star.}
\end{deluxetable}

% Table 3: RV table of eight candidates
\begin{deluxetable}{ccccc}
\tabletypesize{\scriptsize}
\tablenum{3}
\tablecaption{Radial Velocity Measurements of Candidate sdO Components}
\tablewidth{0pt}
\tablehead{
\colhead{HD} & 
\colhead{Date} & 
\colhead{SWP} & 
\colhead{$V_{\rm peak}$} & 
\colhead{$\sigma$}  \\
\colhead{Number} & 
\colhead{(HJD-2400000)} & 
\colhead{Number} & 
\colhead{(km s$^{-1}$)} &
\colhead{(km s$^{-1}$)} }
\startdata
% HD29441
\phn29441&49651.9145&52665&\phn$-$60.7\phs&\phn2.5\\
% HD43544
\phn43544&45698.7251&21911&\phn$-$16.8\phs&\phn4.0\\
% HD51354
\phn51354&45045.3074&16547&\phn13.8&\phn1.8\\
\phn51354&45337.5616&18937&\phn45.5&\phn3.1\\
% HD60855
\phn60855&45698.8371&21915&\phn$-$10.4\phs&\phn3.1\\
\phn60855&47654.3590&36216&\phn\phn2.1&\phn3.8\\
\phn60855&47808.0710&37280&\phn\phn$-$4.0\phs&\phn4.1\\
\phn60855&47908.9317&38036&\phn13.4&\phn2.3\\
\phn60855&48352.5316&41308&\phn\phn1.7&\phn5.9\\
\phn60855&49762.4824&53906&\phn\phn$-$4.2\phs&\phn5.8\\
% HD113120
113120&45602.3331&21155&\phn63.8&\phn3.2\\
113120&46710.9800&29397&\phn$-$14.8\phs&\phn4.0\\
113120&46920.2862&30910&\phn$-$64.1\phs&\phn4.7\\
% HD137387
137387&46225.1106&26120&\phn$-$13.2\phs&\phn2.4\\
137387&46225.3690&26129&\phn$-$12.1\phs&\phn2.0\\
137387&47717.4598&36649&\phn73.1&\phn3.3\\
137387&48346.7955&41240&\phn$-$23.8\phs&\phn3.5\\
% HD152478
152478&47270.2997&33308&\phn$-$42.6\phs&\phn5.5\\
152478&47652.3252&36197&\phn48.3&\phn4.7\\
% HD157042
157042&46527.3127&28114&\phn19.7&\phn5.4\\
157042&46711.9244&29404&\phn25.3&\phn4.5\\
157042&46920.4479&30914&\phn36.2&\phn8.0\\
157042&49442.3921&50427&\phn$-$47.3\phs&\phn6.8\\
% HD157832
157832&49933.1996&55410&\phn15.1&\phn3.3\\
157832&49973.0096&55913&\phn$-$97.6\phs&\phn8.9\\
% HD191610 (25/46 listed)
191610&46245.3612&26301&\phn$-$16.5\phs&\phn2.4\\
191610&46337.6586&26775&\phn10.5&\phn8.9\\
191610&46337.6830&26776&\phn17.6&10.8\\
191610&46337.7275&26778&\phn17.7&\phn7.1\\
191610&46337.7501&26779&\phn11.1&\phn8.9\\
191610&46337.8480&26783&\phn16.0&\phn9.7\\
191610&46337.8987&26785&\phn16.2&15.3\\
191610&46338.7622&26803&\phn19.8&\phn6.5\\
191610&46338.7881&26804&\phn18.4&\phn5.8\\
191610&46338.8322&26806&\phn17.1&\phn9.3\\
191610&46338.8545&26807&\phn11.4&13.3\\
191610&46338.8767&26808&\phn26.0&\phn7.5\\
191610&46692.0815&29241&\phn\phn8.2&\phn5.5\\
191610&47790.2654&37096&\phn10.7&16.9\\
191610&47790.3374&37098&\phn\phn8.5&\phn4.7\\
191610&47791.2331&37124&\phn15.9&10.6\\
191610&47791.3635&37128&\phn21.3&16.4\\
191610&47791.4270&37130&\phn\phn3.5&11.8\\
191610&47791.5566&37134&\phn14.5&\phn5.8\\
191610&47791.6179&37136&\phn\phn8.0&13.4\\
191610&47791.7448&37140&\phn\phn3.3&12.1\\
191610&47791.8080&37142&\phn12.4&14.7\\
191610&47792.0716&37148&\phn\phn8.4&16.5\\
191610&47792.2865&37150&\phn18.2&\phn6.5\\
191610&47792.4370&37154&\phn\phn6.7&\phn5.9\\
% HD194335
194335&45463.4591&19938&\phn46.3&\phn6.5\\
194335&49349.8095&49705&\phn$-$79.4\phs&\phn4.0\\
194335&49470.4343&50637&\phn$-$96.1\phs&\phn9.3\\
194335&49686.9274&52946&\phn29.5&\phn3.9\\
% HD214168 (9/20 listed)
214168&47691.1849&36479&$-$118.4\phs&\phn5.6\\
214168&47769.1105&36901&\phn$-$32.9\phs&\phn2.3\\
214168&49296.6583&49099&\phn\phn2.3&\phn3.6\\
214168&49296.7311&49102&\phn\phn8.3&\phn7.3\\
214168&49296.7780&49104&\phn\phn2.4&\phn6.5\\
214168&49298.6622&49130&\phn\phn3.0&\phn2.8\\
214168&49298.7078&49132&\phn21.2&\phn5.3\\
214168&49299.6744&49145&\phn23.4&\phn5.8\\
214168&49299.7211&49147&\phn21.6&\phn5.0
\enddata
\end{deluxetable}

%%%%%%%%%%%%%%%%%%%%% Figures

% Figure 1: CCFs plots of Secondary, Uncertain, Primary correlations
\begin{figure}
\begin{center} 
{\includegraphics[height=12cm]{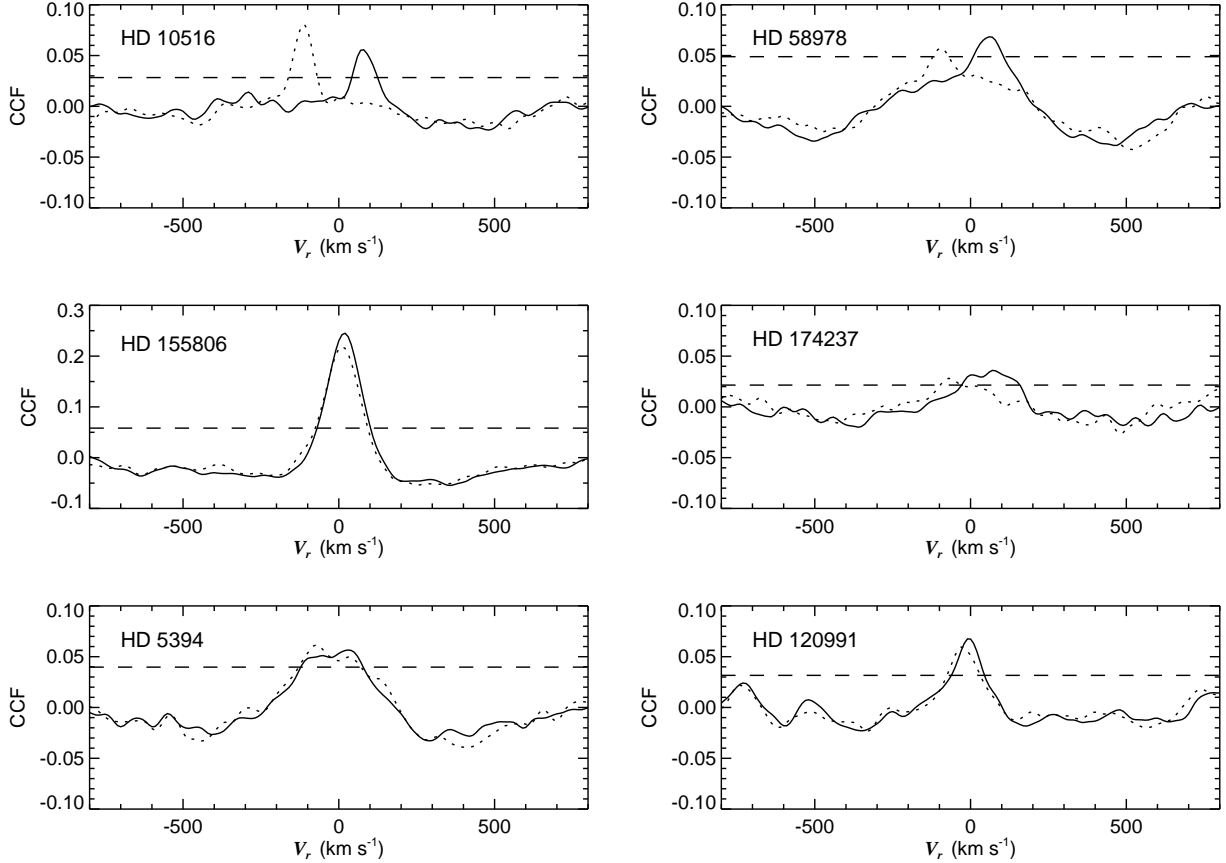}} 
\end{center} 
\caption{Example CCFs for two cases of hot component detection (top row) 
with several other cases where the peak is associated with the Be component 
(second and third rows). The top panels show the CCFs for HD 10516, a confirmed 
system with a hot subdwarf companion \citep{gies1998}, and HD 58978, a confirmed 
system with a fainter subdwarf companion \citep{peters2008}. 
The middle panels show how the spectrum of the emission-line star contributes 
more to the CCF for a hot star (HD 155806; O7.5 \rom{3}e) 
than a mid-range temperature star (HD 174237; B4 \rom{4}(e)). 
The lower row illustrates how the CCF from the 
rapidly rotating Be component is usually very broad 
(HD 5394; $V\sin{i} = 432$ km s$^{-1}$), but sometimes narrow 
(HD 120991; $V\sin{i} = 70$ km s$^{-1}$). 
Examples of the CCFs for blue and red Doppler-shifted spectra are plotted as 
dotted and solid lines, respectively. The horizontal dashed lines indicate the 
S/N=3 limit for detection.} 
\label{CCFs plot of non-candidates}
\end{figure}

% Figure 2: CCFs plots of eight candidates
\begin{figure} 
\begin{center} 
{\includegraphics[height=14cm]{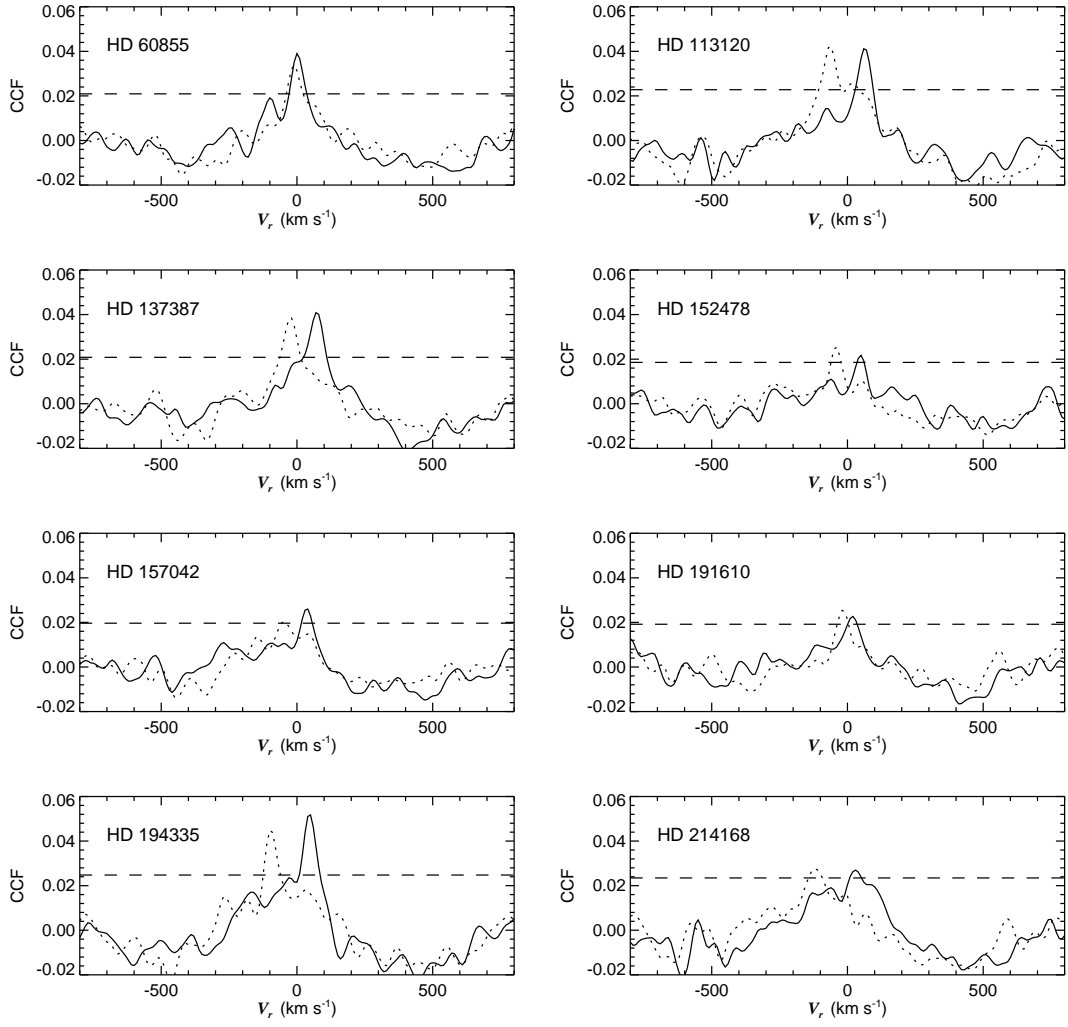}} 
\end{center} 
\caption{CCF plots of eight Be+sdO binary candidates in the same format as Fig.~1.} 
\label{CCFs plot of eight candidates}
\end{figure}   

% Figure 3: CCFs plot of four C? 
\begin{figure}
\begin{center}
{\includegraphics[height=7cm]{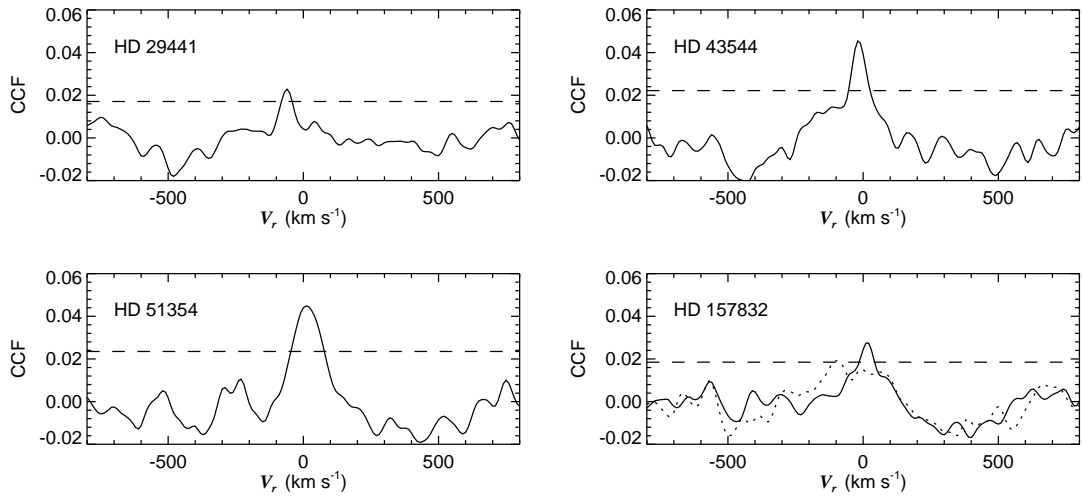}}
\end{center}
\caption{CCF plots of four potential Be+sdO binary candidates in the same format as Fig.~1.}
\label{CCFs plot of C?}
\end{figure}
 
% Figure 4: Histogram of Spectral Type Distribution
\begin{figure}
\begin{center}
{\includegraphics[height=10cm]{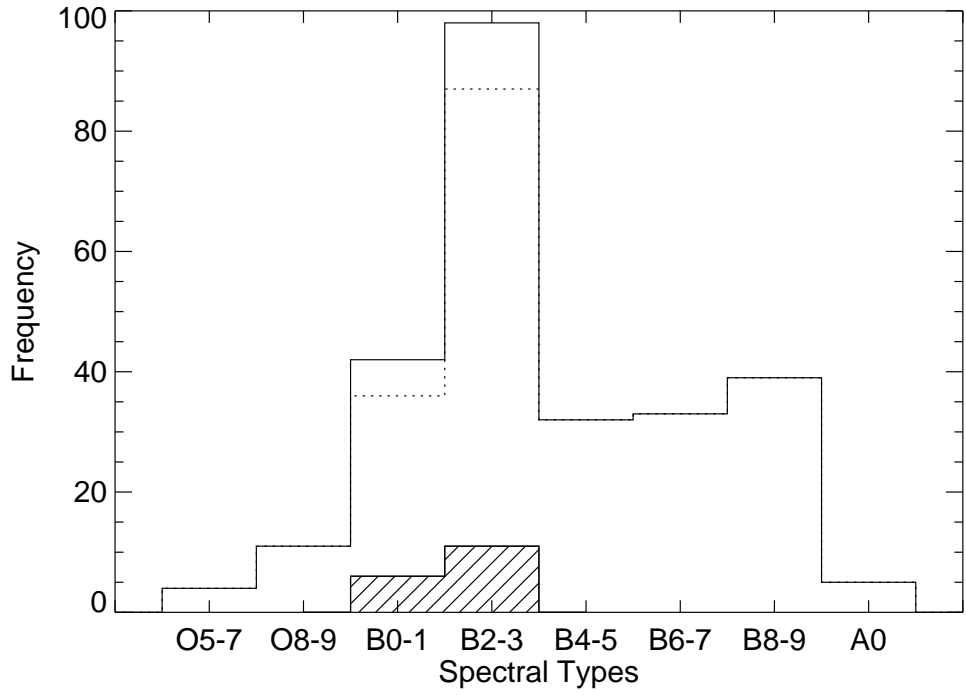}}
\end{center}
\caption{Histograms of the spectral type distributions of the full sample (solid line),
those with no detections (dotted line), and those with known or candidate Be+sdO systems 
(line filled).}
\label{SpT_Hist}
\end{figure}

\end{document}